\documentclass[slac_one]{revtex4}
\usepackage{graphicx}
\usepackage{fancyhdr}
\begin{document}

\title{Study of VV Scattering Processes as a Probe of Electroweak Symmetry Breaking}

\author{A. Sznajder, on behalf of the CMS Collaboration}
\affiliation{UERJ, Rio de Janeiro, RJ 20550-900, Brazil}

\begin{abstract}
An exploratory study has been performed in order to assess the possibility of probing the symmetry breaking mechanism through the $VV$ fusion process using the CMS detector. A model independent analysis was carried out with no assumption on the mechanism restoring the unitarity in the scattering amplitude and without any degrees of freedom beyond the SM. In order to explore the sensitivity of the analysis method to an heavy Higgs resonance, we analyzed a data set produced using an Higgs boson mass of $500GeV$. Moreover, in order to consider the $VV$ fusion cross section in a region where no resonances are present, a sample corresponding to the no-Higgs scenario, that in the SM is equivalent to a very high Higgs mass, has been also studied.
\end{abstract}

\maketitle

\thispagestyle{fancy}

\section{INTRODUCTION}

Vector boson scattering is the key process to probe the nature of the electroweak symmetry breaking (EWSB) mechanism. In absence of the Higgs particle, the Standard Model (SM) predicts that the scattering amplitude of longitudinally polarized vector boson ($V_LV_L\rightarrow V_LV_L$) grows linearly with $s$, violating unitarity at about $1.5~TeV$. This implies that the SM becomes a strongly interacting theory at high energy, so one is led to expect the presence of resonances in the boson-boson fusion process. Alternatively, new physics must intervene below the $TeV$ scale. Several theories have been proposed, such Little Higgs, Dynamical symmetry breaking and  Higgless models.

On the other hand, if a massive Higgs boson exists, a resonance will be observed in the $VV$ invariant mass distribution. In fact the $VV$ fusion process represents the second most important contribution to Higgs production at LHC. 

The study \cite{cmsnote} has been carried on several processes with at least one vector boson decaying into leptons:
$qq \rightarrow qqVV \rightarrow qqVZ \rightarrow qqqq\mu\mu / ee$,
$qq \rightarrow qqVV \rightarrow qqVW \rightarrow qqqq\mu\nu / e\nu$,
$qq \rightarrow qqVV \rightarrow qqZZ \rightarrow qq\mu\mu\mu\mu / qqeeee$,
$qq \rightarrow qqVV \rightarrow qqZW \rightarrow qq\mu\mu\mu\nu$,
$qq \rightarrow qqVV \rightarrow qqW^{\pm} W^{\pm} \rightarrow qq\mu^{\pm}\nu\mu^{\pm}\nu$.

\section{SIGNAL AND BACKGROUND}

\subsection{Signal Simulation}

To access the sensitivity for new physics of the $VV$ scattering, a Monte Carlo study \cite{cmsnote} has been made, using two scenarios: the no-Higgs and Higgs (m$_H$=500~GeV). The generated events have been processed through the CMS detector simulation which includes low luminosity pile-up.

The signal generation is a key aspect of the study, since a precise knowledge of the $\sigma(pp\rightarrow VVjj)$ on the whole $VV$ invariant mass spectrum is essential.

The decay of the vector bosons produces a six fermion final state. When taking into account the finite width of the EW bosons and simulating off-shell vector bosons, the full set of diagrams describing $qq\rightarrow 6~\mathrm{fermions}$ has to be considered in order to take into account the quantum interferences. For this purpose, the $qq\rightarrow 6 ~\mathrm{fermions}$ process has been simulated using the $PHANTOM$ ~\cite{phase_art_1,phase_art_2,phase_art_3} matrix element generator$(\alpha_{EW}^{6})$.

In order to investigate EWSB, the $VV$ fusion processes are isolated from all the others six fermion final states, the irreducible backgrounds, by applying kinematic selections which enhances regions of phase space dominated by the signal.

\subsection{Backgrounds}

Besides the irreducible background, which is determined as the complementary of the signal, the most problematic background for the VBF signal in the semileptonic final state is the production of a single $W$ (or $Z$) in association with jets, while for the totally leptonic channel is instead the production of a pair of bosons in association with jets. A potentially dangerous background, because of its big cross section, is the QCD production of top pairs.

\section{SIGNAL RECONSTRUCTION AND BACKGROUND REJECTION}

The jets are considered as fermion candidates in case they are not identified as b-jets and have a minimal p$_T$ of 30~GeV/c, while leptons are required to satisfy basic quality selections for the identification.

The particles in the final state coming from the decay of a $W$ (or a $Z$) are expected to have high transverse momentum ($p_T$) and to be mostly produced centrally in the detector.

\begin{figure}[!Hhb]
\begin{center}
\includegraphics[width=.4\textwidth]{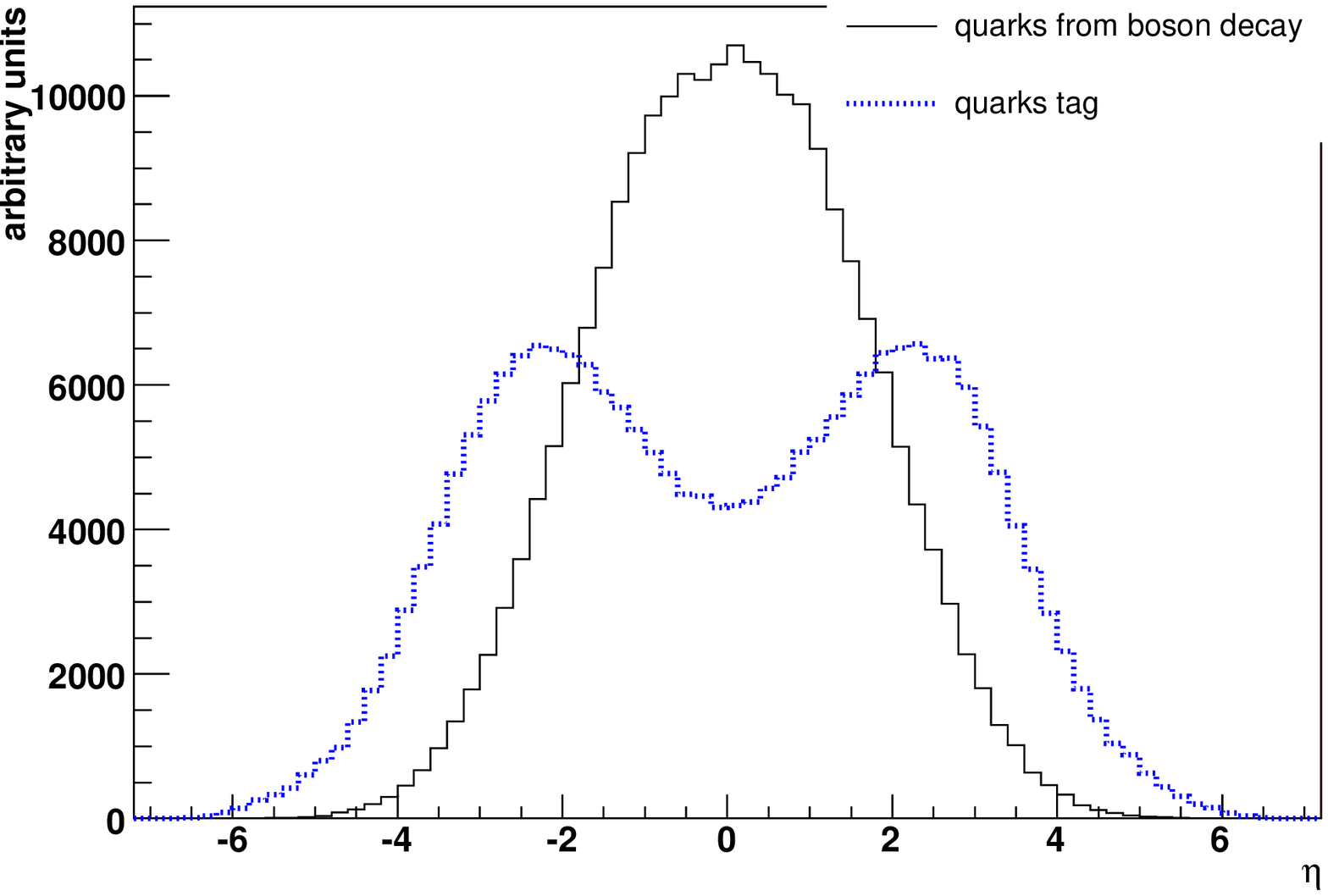}
\includegraphics[width=.4\textwidth]{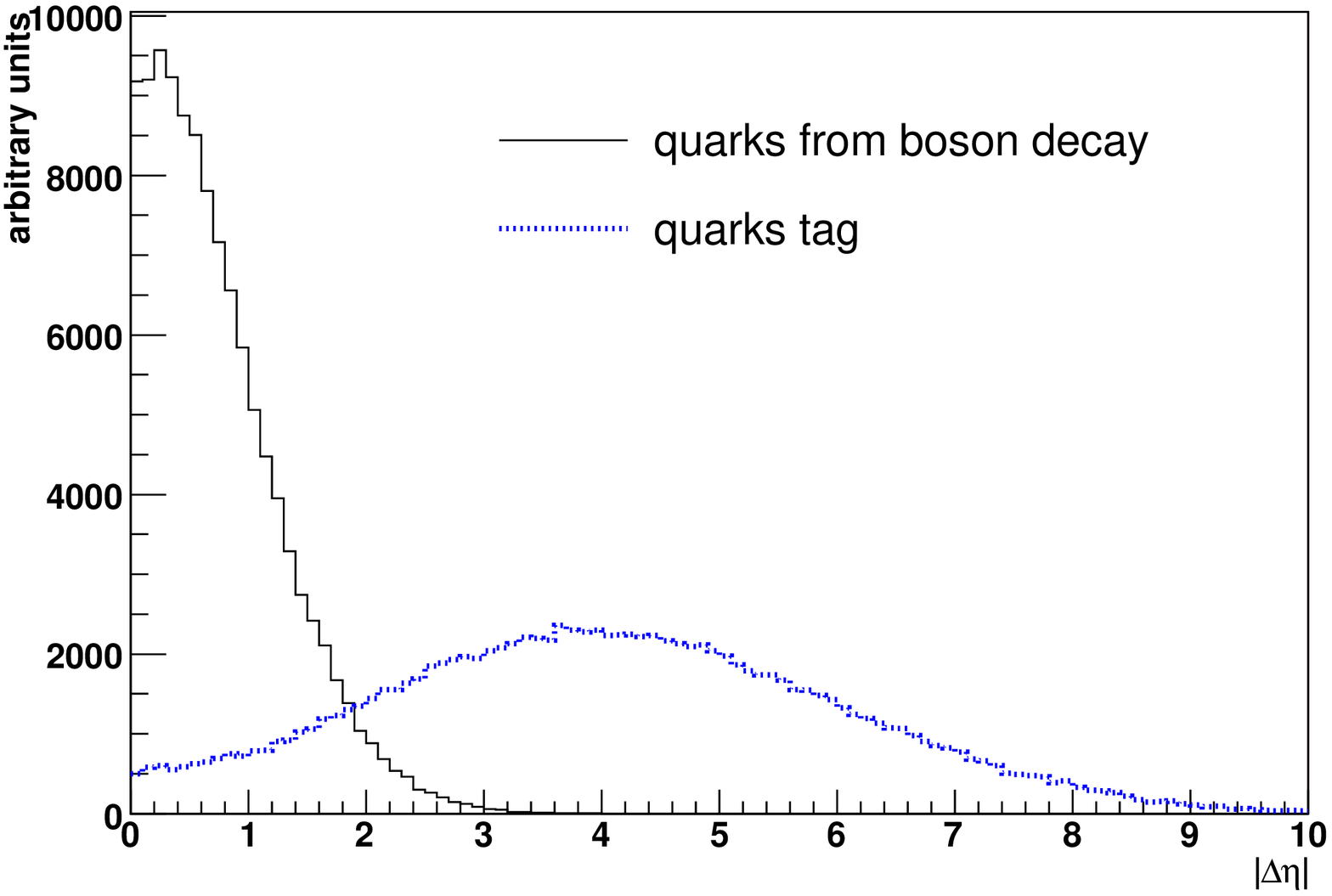}
\caption{Pseudorapidity of the two quarks from the boson decay(left) and distance in pseudorapidity of the two quarks from the boson decay and of the two tag quarks in the no-Higgs scenario for the $qqqq\mu\nu$ final state.}
\label{fig:cinem_quarks}
\end{center}
\end{figure}

On the contrary, the two quarks that have radiated the vector bosons, the tag quarks, tend to go in the forward/backward regions at high $\vert\eta\vert$ and they have very large energy and $p_T$. In Fig.\ref{fig:cinem_quarks} the $\eta$ of the two quarks from the boson decay and of the two tag quarks are compared.

%\begin{figure}[!Hhtb]

This kinematic pattern is used to tag the VBF events as a six fermions final state: the tag jets are identified as the pair of jets with the highest invariant mass and for the signal they are required to have a minimal separation ($\vert\Delta\eta\vert\:>\:1.5$) and a minimal invariant mass of $500~GeV/c^{2}$.

Moreover, a set of kinematic cuts has been applied in order to distinguish the signal events from all the processes considered as backgrounds. The jets produced by a vector boson are selected as the ones with minimal $\vert\Delta\eta\vert$. The leptons are required either to be coupled as products of a $Z$ decay, or to be identified in an event with a minimal measured transverse missing energy in the $W$ case. Once the vector bosons are identified, their sum is required to have at least 100~GeV/c$^{2}$ invariant mass.

\begin{figure}[!Hhb]
\begin{center}
\includegraphics[width=.4\textwidth]{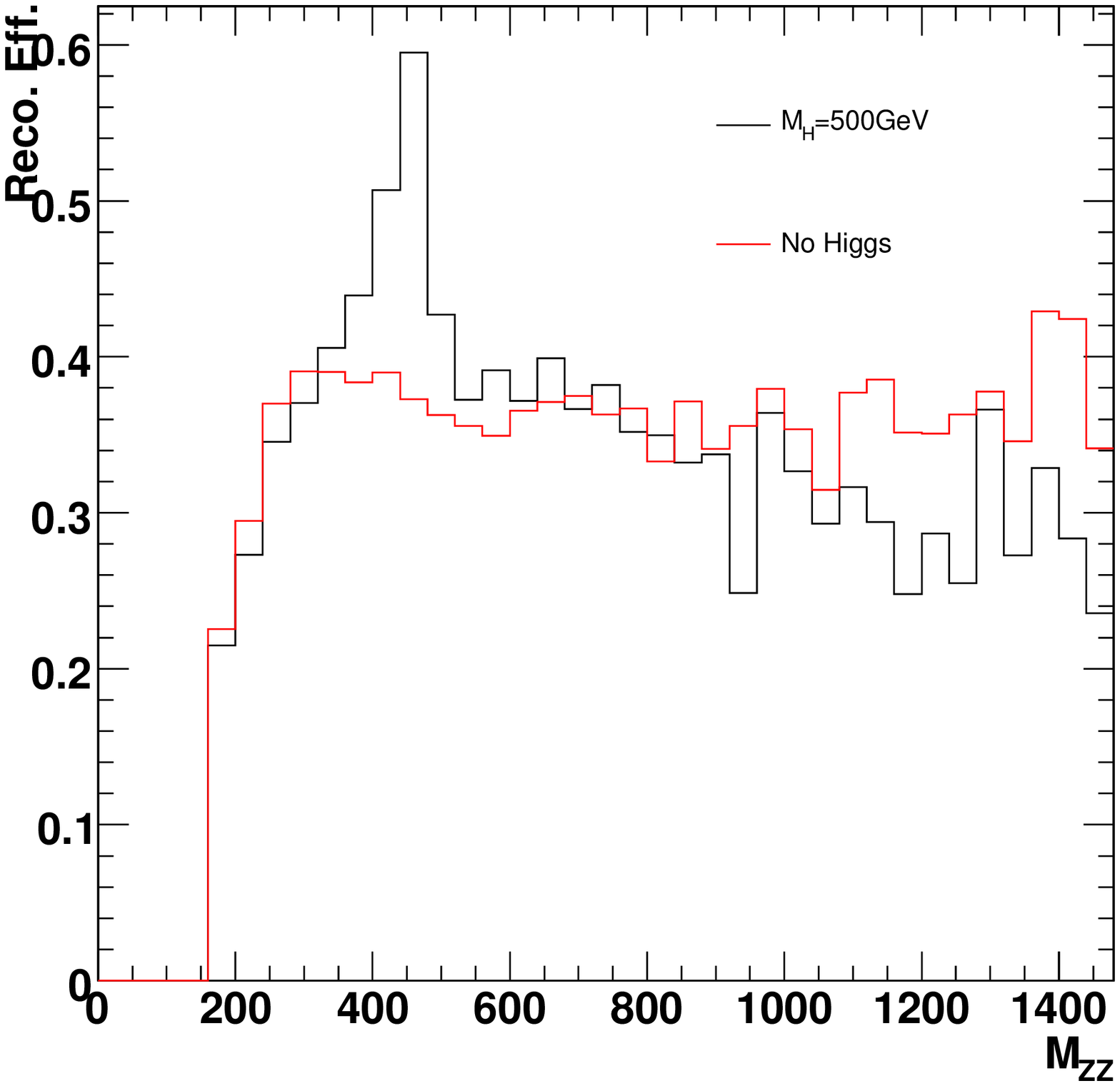}
\includegraphics[width=.4\textwidth]{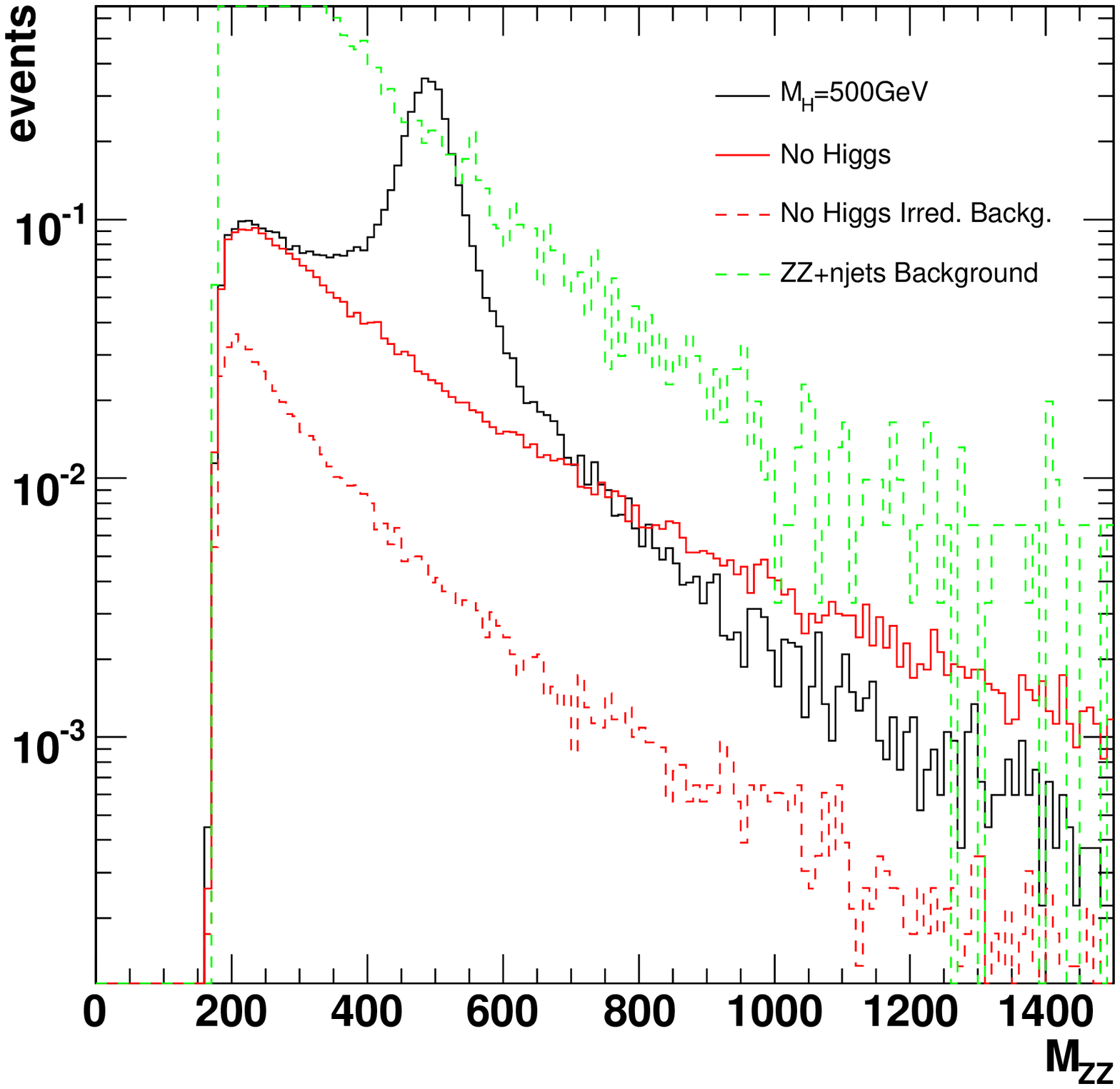}
\caption{Efficiency of the signal reconstruction and signal and background spectra, for Higgs and no Higgs scenarios, for the $qq\mu\mu\mu\mu$ final state.}
\label{fig:eff}
\end{center}
\end{figure}

As an example, Fig. \ref{fig:eff} shows the efficiency of the signal reconstruction, for both scenarios, and the  signal and background spectrums for the $qq\mu\mu\mu\mu$ channel.

%\begin{figure}[!Hhbt]

For the $500~GeV$ Higgs scenario, the expected number of signal and background events as well as the significances for all studied channels, after $60~fb^{-1}$ of integrated luminosity, is reported in Tab. \ref{tab:evnum}.

\begin{table}[!Hhbt]
\begin{tabular}{|l|c|c|c|c|c|c|c|c|}  
\hline
                        & $qqqq\mu\nu$ & $qqqqe\nu$ & $qqqq\mu\mu$ & $qqqqee$ & $qq\mu\mu\mu\mu$ &  $qqeeee$\\  
\hline
\bfseries{signal}       & 703          &    309      &      86    &     100     &        3.1       &   3.5           \\
\hline
\hline
$W$ + n jets            & 34840       &    1383      &     -      &          -  &        -         &      -          \\
\hline
$Z$ + n jets            & 3094        &    -         &     3798   &      4660   &        -         &      -          \\
\hline
$\bar{t}t$              & 5976        &    609       &     30     &      14     &        0         &      -          \\ 
\hline
$ZZ$ + n jets           & -           &    -         &     125    &      184    &        2.6       &    2.9     \\
\hline
$ZW$ + n jets           &  -          &    -         &     781    &        615  &         0        &     -          \\
\hline  
$WW$ + n jets           & 16133       &    -         &      -     &        -    &         0        &     -          \\
\hline    
irreducible backgrounds & 220         &    23      &      20     &        20   &        0.036      &    0.04          \\
\hline
\hline
\bfseries{backgrounds}  & 60263     &    2015      &   4754      &       5493  &        2.6       &    2.94          \\
\hline
\hline
\bfseries{significance} & 2.86       &    6.72      &   1.24     &      1.34     &        1.66      &    1.76         \\
\hline

\end{tabular}

\caption{Number of signal and backgrounds events after the analysis cuts for each final state with $60~fb^{-1}$ of integrated luminosity}

\label{tab:evnum}

\end{table}

\section{CONCLUSIONS}

An exploratory study of the VV-scattering process at CMS as probe of Electroweak Symmetry Breaking is reported. Both in the Higgs and in the no-Higgs scenario, either a peak or a deviation for the Standard Model predicitions will be observed in the M($VV$) spectrum. The $VV$ scattering holds then the key to understand the Electronweak Simmetry Breaking, in a model independent way.

\end{document}